\documentclass[article,reprint,amsmath,amssymb,superscriptaddress,longbibliography]{revtex4-1}
\usepackage{graphicx}
\usepackage{dcolumn}
\usepackage{bm}

\usepackage[colorlinks,
            linkcolor=blue,
            anchorcolor=blue,
            citecolor=blue,
            ]{hyperref}
\usepackage{breakurl}
\usepackage{multirow}
\usepackage{enumitem}
\usepackage{tablefootnote}
\usepackage{threeparttable}
\usepackage{tabularx}
\setcitestyle{super} 
\usepackage{lineno}

\begin{document}
	

 \title {Nonvolatile electric switching of critical current in cross-bar superconducting junctions}
	

   \author{Jiajun Ma}
   \affiliation{Key Laboratory for Quantum Materials of Zhejiang Province, Department of Physics, School of Science and Research Center for Industries of the Future, Westlake University, Hangzhou 310030, P. R. China}
   \affiliation{Institute of Natural Sciences, Westlake Institute for Advanced Study, Hangzhou 310024, P. R. China}
   
   \author{Jingyi He}
   \affiliation{Key Laboratory for Quantum Materials of Zhejiang Province, Department of Physics, School of Science and Research Center for Industries of the Future, Westlake University, Hangzhou 310030, P. R. China}
   \affiliation{Institute of Natural Sciences, Westlake Institute for Advanced Study, Hangzhou 310024, P. R. China}

   \author{Qiong Qin}
   \affiliation{Key Laboratory for Quantum Materials of Zhejiang Province, Department of Physics, School of Science and Research Center for Industries of the Future, Westlake University, Hangzhou 310030, P. R. China}
   \affiliation{Institute of Natural Sciences, Westlake Institute for Advanced Study, Hangzhou 310024, P. R. China}

   \author{Tian Le}
   \affiliation{Center for Quantum Matter, School of Physics, Zhejiang University, Hangzhou 310058, China}


   \author{Zhiwei Wang}
   \affiliation{Centre for Quantum Physics, Key Laboratory of Advanced Optoelectronic Quantum Architecture and Measurement (MOE), School of Physics, Beijing Institute of Technology, Beijing 100081, China}
   \affiliation{Beijing Key Lab of Nanophotonics and Ultrafine Optoelectronic Systems, Beijing Institute of Technology, Beijing 100081, China}
   \affiliation{Beijing Institute of Technology, Zhuhai 519000, China}
   
   
   \author{Jie Wu}
   \affiliation{Key Laboratory for Quantum Materials of Zhejiang Province, Department of Physics, School of Science and Research Center for Industries of the Future, Westlake University, Hangzhou 310030, P. R. China}
   \affiliation{Institute of Natural Sciences, Westlake Institute for Advanced Study, Hangzhou 310024, P. R. China} 
   
   \author{Congjun Wu}
   \affiliation{Key Laboratory for Quantum Materials of Zhejiang Province, Department of Physics, School of Science and Research Center for Industries of the Future, Westlake University, Hangzhou 310030, P. R. China}
   \affiliation{Institute of Natural Sciences, Westlake Institute for Advanced Study, Hangzhou 310024, P. R. China} 
   
   \author{Xiao Lin}
   \affiliation{Key Laboratory for Quantum Materials of Zhejiang Province, Department of Physics, School of Science and Research Center for Industries of the Future, Westlake University, Hangzhou 310030, P. R. China}
   \affiliation{Institute of Natural Sciences, Westlake Institute for Advanced Study, Hangzhou 310024, P. R. China}
   
   \date{\today}
	
\begin{abstract}
	\noindent
	Superconducting (SC) diodes are key passive  building blocks for future SC electronics. However, realizing their active counterparts is essential for functional logic. Here, we demonstrate deterministic nonvolatile electrical switching of the critical current ($I_\text{c}$) in overlap crossbar SC junctions.  By applying a minimal perpendicular magnetic field ($H_\text{z}$), $I_\text{c}$ is modulated by a factor of four with a large switching efficiency of 60\%, achieved at a significantly reduced excitation current density of $5\times10^5$~A/cm$^2$. We also uncover anomalous behaviors: an electrically switchable critical temperature and a non-monotonic $I_\text{c}$-$H_\textit{z}$ response. These observations are interpreted in terms of unique asymmetry involving isolated vortex injection, configuration and repulsion inherent to the junction geometry.  
	Our device provides a scalable, low-power alternative to complex SQUID-based architectures, paving the way for high-density SC integrated circuits.


	
\end{abstract}

\maketitle

\begin{figure*}[thb]
	\includegraphics[width=0.7\textwidth]{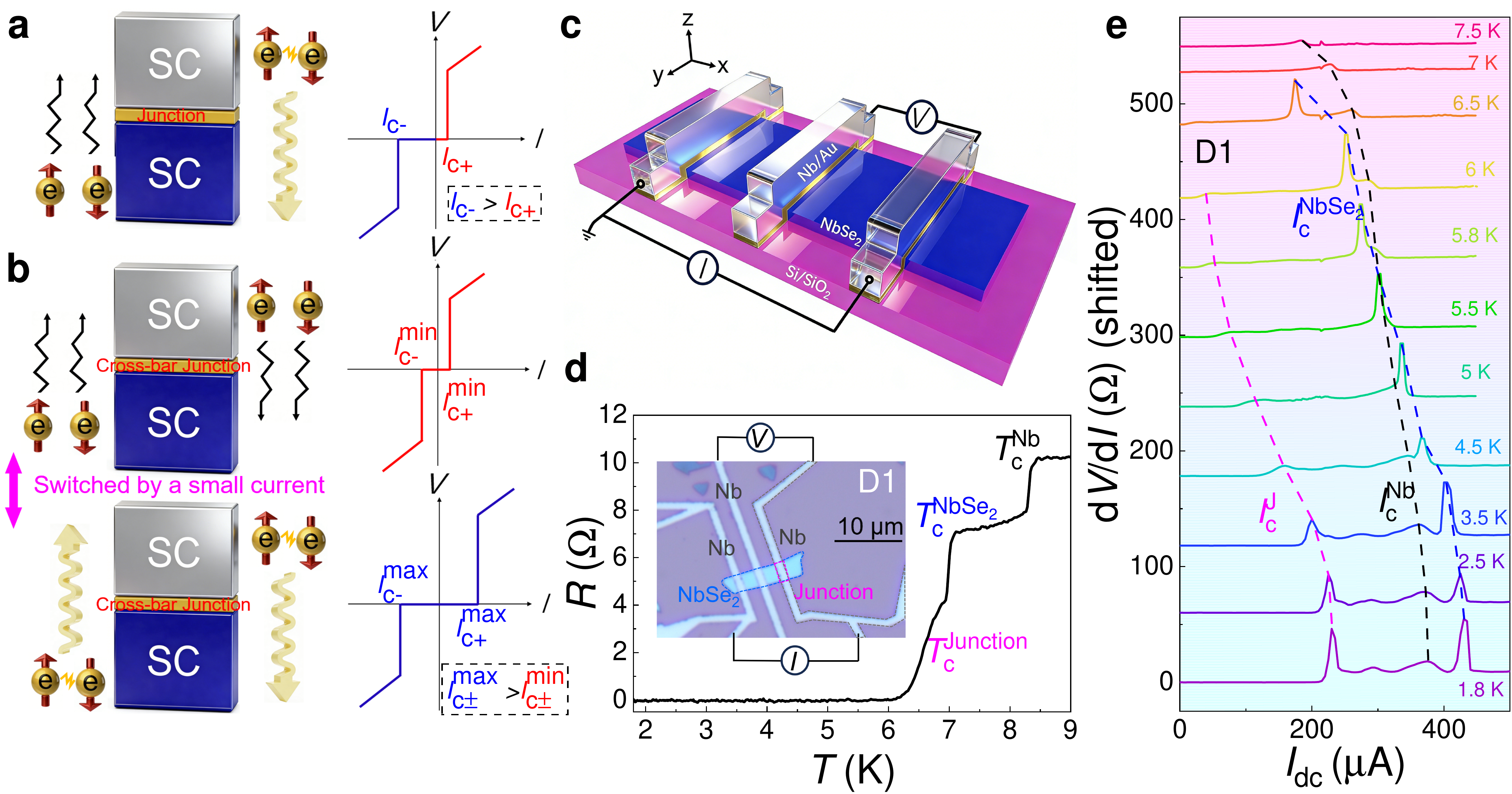}
	\caption{\textbf{Basis transport properties of D1.} \textbf{a} Illustration of SDEs, exhibiting critical current  asymmetry. \textit{b} Illustration of switchable SC devices, showing critical current manipulation via current excitation.  		
		\textbf{c} Schematic illustration of SC junction made of NbSe$_2$/Au/Nb. \textbf{d}	$T$-dependence of the resistance for D1. The inset is the optical image of the device. \textbf{e} d$V$/d$I$ as a function of $I_\textrm{dc}$ measured at various $T$ for D1. The dashed curves mark three transition peaks of NbSe$_2$, Nb and the junction, respectively. Scale bar, 10 $\mu$m.}
		\label{Fig1}
\end{figure*}

\noindent
Superconducting electronics holds immense promise as a next-generation electronic technology with ultra-high energy efficiency and speed, offering transformative potential in SC information processing and storage~\cite{Likharev1979IEEE,Aziz2023NE}. 
SC diode effect (SDEs),  a dissipationless analogue to nonreciprocal charge transport in semiconducting diodes, exhibit SC state under forward bias current and resistive  state under reverse bias , enabling rectification~\cite{HuJP2007PRL,WangXL2023NRP,MaJJ2025APR}.  
Following its initial observation in [Nb/V/Ta]$_\textrm{n}$ superlattices~\cite{Ono2020Nature}, SDEs have been observed in a variety of SC systems, either in intrinsic material~\cite{LeT2024Nature,DuanXF2024Nature,WangJ2025NC,Paradiso2022NC}, heterostructure~\cite{Ono2020Nature,Ono2022NN} or and Josephson junctions (JJs, seen in Fig.~\ref{Fig1}a) ~\cite{Parkin2022NM,Ali2022Nature,Franke2023Nature,Deshmukh2024NM,Krasnov2022NC,ShenJ2024NC,WangKL2025NC}. The ability of SDE to allow asymmetric supercurrent flow make it highly attractive for applications in passive SC circuits and potentially for energy harvesting \cite{XuHQ2024PRL,LiaoZM2025Arxiv,ZhangD2025Arxiv}.



Consequently, the development of switchable SC electronic devices is of paramount importance. Such devices would serve as a fundamental building block for programmable logic gates and memories~\cite{Aziz2023NE}. In this context, conventional superconductor/insulator/superconductor (SIS) JJs and SQUID designs have received long-term attention, however their poor scalability poses a major  bottleneck for highly integrated application~\cite{Aziz2023NE,Nagasawa2026SST,Polyakov2007IEEE,Jackman2017IEEE}. Alternative architectures, such as planar JJs and ferroelectric/SC heterostructures, offer the potential for non-volatile electric switching of the critical current ($I_\text{c}$)~\cite{Krasnov2015NC,Ivry2021APL}. Parallel efforts on ferromagnet/superconductor/ferromagnet (FM/S/FM) heterostructures have achieved critical temperature ($T_\textit{c}$) modulation via  external magnetic field ($H$), a phenomenon known as the SC spin valve (SSV) effect~\cite{You2002PRL,Blamire2017NM,Robinson2025NC}. However, these approaches face specific limitations, ranging from fabrication complexity, poor scalability, large $H$ (kGs),  considerable activation current (mA) to long switch time (s). 


In this work, we present an alternative design of switchable SC electronics, leveraging the facile fabrication of sandwiched crossbar NbSe$_2$/Au/Nb junctions (Fig.~\ref{Fig1}b, c). The device demonstrates significant nonvolatile control of both $I_\text{c}$ and $T_\textit{c}$ via moderate electric activation under a minimal perpendicular field ($H_\text{z}$) down to 0.1~Gs. 
A maximal fourfold modulation of $I_\text{c}$ is achieved with an activation current on the order of $I_\text{c}$, establishing a substantial window for operation. The calculated activation current density amounts to $5\times10^5$~A/cm$^2$, orders of magnitude lower than those of other designs \cite{Krasnov2015NC,MiaoF2024NC}. Deterministic, nonvolatile switching between the SC and normal states is demonstrated using short current pulses (50 $\mathrm{\mu}$s). Beyond this, we observe an unusual non-monotonic dependence of $I_\text{c}$ on $H_\textit{z}$ characterized by discontinuous jumps. {\color{black}We attribute these phenomena to intricate vortex dynamics, specially associated with the crossbar architecture, involving asymmetric Abrikosov vortex injection, configuration and repulsion.} 
This architecture, with its minimal operating requirements and excellent suitability to miniaturization, unlocks a novel pathway toward next-generation cryogenic logic and memory devices.

~\\
\noindent{\Large\textbf{Results}}

\noindent\textbf{Electric switchable critical current}

\noindent
{\color{black}Figure~\ref{Fig1}} characterizes the basic transport properties of device 1 (D1). The temperature ($T$) dependent resistance ($R$) in {\color{black}Fig.~\ref{Fig1}d} reveals three distinct SC transitions at 8.3~K, 7~K and 6.5~K, attributed to Nb, NbSe$_2$, and the junction, respectively. {\color{black}$T_\text{c}$ of Nb and NbSe$_2$ match the bulk value, underscoring high device quality.} The differential resistance ($dV/dI$) versus bias current ($I_\text{dc}$) is plotted in {\color{black}Fig.~\ref{Fig1}} at various $T$. The critical currents $I_\textrm{c}^\mathrm{NbSe_2}$, $I_\textrm{c}^\mathrm{Nb}$ and $I_\textrm{c}^\mathrm{J}$ are clearly identified, each approaching zero by increasing $T$ towards the corresponding $T_\textrm{c}$, as traced by dashed curves.

\begin{figure*}[thb]
	\begin{center}
		\includegraphics[width=0.8\textwidth]{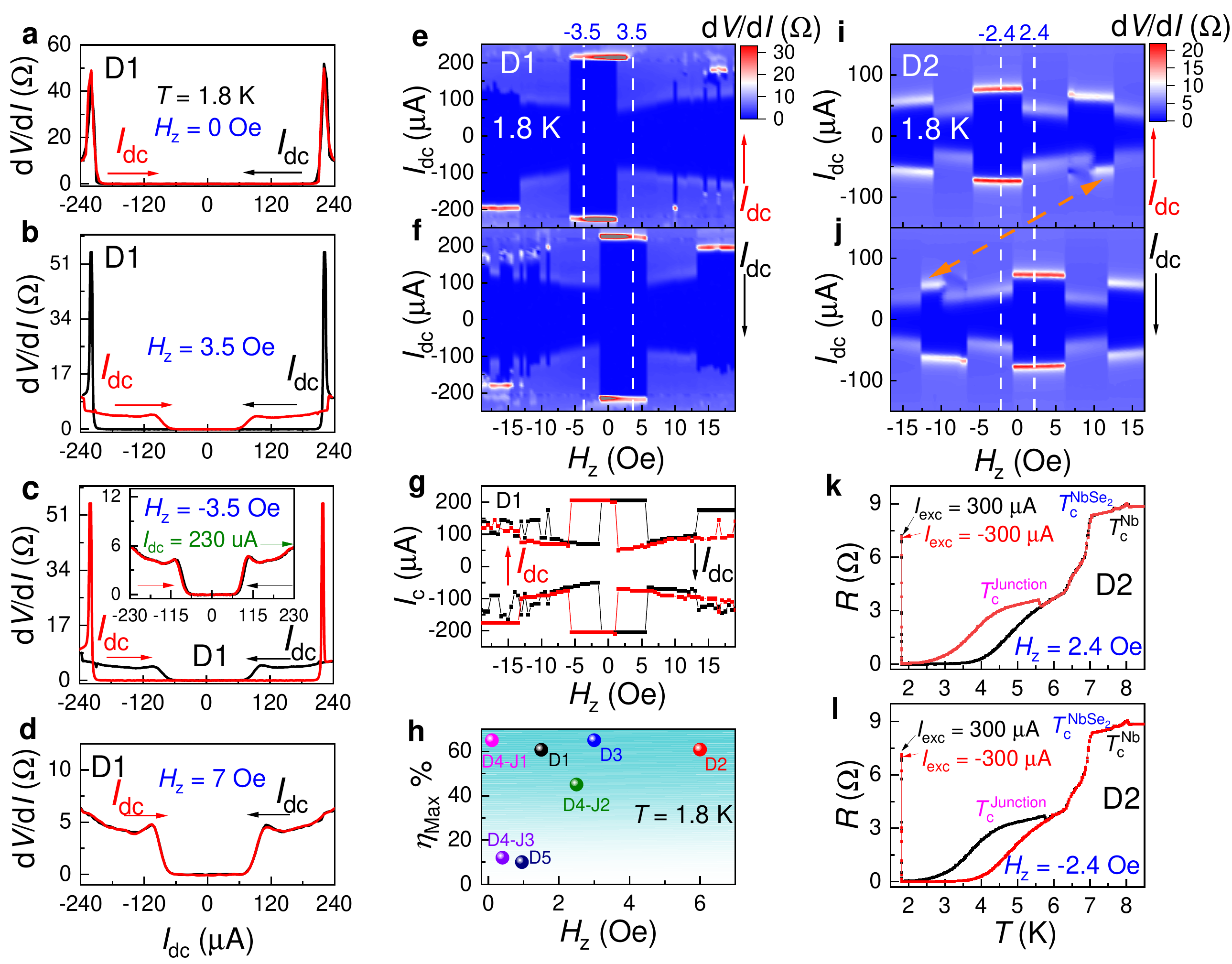}
	\end{center}
	\setlength{\abovecaptionskip}{-8 pt}
	\caption{\textbf{Electric switching characteristics in D1 and D2}. \textbf{a-d} d$V$/d$I$ versus $I_\textrm{dc}$ with a maximum bias of 240~$\mu$A at selective $H_\textit{z}$ for D1. The inset of \textbf{c} exhibits d$V$/d$I$ versus $I_\textrm{dc}$ with  a maximum bias of 230~$\mu$A at  $H_\textit{z}=-3.5$~Oe. The red and black curves denote forward and backward bias sweeps, respectively. \textbf{e-f}  d$V$/d$I$ maps as a function of $I_\textrm{dc}$ and $H_\textit{z}$ for D1. \textbf{g} Extracted $I_\text{c}^\text{J}$ as a function of $H_\text{z}$ for D1. The red and black curves denote forward and backward bias sweeps, respectively. \textbf{h} Summary of switching efficiency ($\eta$) across different devices. \textbf{i-j}  d$V$/d$I$ maps as a function of $I_\textrm{dc}$ and $H_\textit{z}$ for D2. \textbf{k-l} $R-T$ profiles of D2 switched by current pulse ($I_\text{exc}=\pm300~\mu$A) at $H_\text{z}=\pm2.4$~Oe, respectively. 
	}
	\label{Fig2}
\end{figure*}

$dV/dI$ versus $I_\text{dc}$ under selective $H_\text{z}$ is presented in {\color{black}Fig.~\ref{Fig2}} for D1. In {\color{black}Fig.~\ref{Fig2}a} and {\color{black}d}, the curves obtained during forward ($-\rightarrow+$) and backward ($+\rightarrow-$) bias sweeps overlap at $H_\text{z}=0$ and 7~Oe, featuring overdamped SC junctions. The maximum bias was set to 240~$\mu$A in order to focus on the junction transition ($I_\mathrm{c}^\text{J}\approx200~\mu$A). In stark contrast, at an intermediate field of $H_\text{z}=3.5$~Oe in {\color{black}Fig~\ref{Fig2}b}, pronounced hysteresis emerges with distinct $I_\mathrm{c}^\text{J}$ observed during forward (red) and backward (black) sweeps. Reversing the field inverts this asymmetry in  {\color{black}Fig~\ref{Fig2}c}. This hysteresis is unlikely to stem from Joule heating or re-trapping effect~\cite{Barone1982book}, since it doesn't appear in {\color{black}Fig.~\ref{Fig2}a and d}. Close inspection of the data reveals $I_\mathrm{c+}^\text{J}=I_\mathrm{c-}^\text{J}$ within individual sweep, exhibiting no SDEs.  

In the inset of {\color{black}Fig.~\ref{Fig2}c}, reducing the maximum bias eliminates the hysteresis,  pointing to the existence of a threshold activation current ($I_\mathrm{sh}$) required to trigger the switching dynamics ($230~\mu\mathrm{A} < I_\mathrm{sh} < 240~\mu\mathrm{A}$). $I_\mathrm{sh}$ is an order of magnitude lower than that of other switchable SC devices, including planar Josephson junctions~\cite{Krasnov2015NC} and magnetic/superconducting van der Waals heterostructures~\cite{MiaoF2024NC}. The corresponding threshold current density ($J_\text{sh}$) approximately {\color{black}amounts to $5\times10^5$~A/cm$^2$}, orders of magnitude lower than the state of art value for switchable spintronic devices~\cite{Manchon2019RMP,Nakatsuji2020Nature,Nakatsuji2022Nature,MiaoF2025PRL,Du2025PRL}. Crucially, the ratio $I_\mathrm{sh}/I_\mathrm{c}$ is close to unity, suggesting a pathway toward highly efficient, low-power switching operations.


To provide further insight, we present color maps of $dV/dI$ as a function of $I_\text{dc}$ and $H_\text{z}$ in {\color{black}Fig.~\ref{Fig2}e} and {\color{black}f} for both bias sweep direction. The extracted $I_\mathrm{c}^\text{J}$ are plotted in {\color{black}Fig.~\ref{Fig2}g}. Remarkably, $I_\mathrm{c}^\text{J}$ showcases sudden jumps as $H_\text{z}$ varies. Specially in {\color{black}Fig.~\ref{Fig2}i and j}, $I_\mathrm{c}^\text{J}$ exhibits a non-monotonic staggered manner by varying $H_\text{z}$ for D2. Within the figures, $I_\mathrm{c}^\text{J}$ is symmetric in positive and negative bias regimes within any individual sweep. But it differs between the opposite sweeps as highlighted by vertical lines in {\color{black}Fig.~\ref{Fig2}e,f} and {\color{black}2i,j}, which underpins the hysteresis observed in {\color{black}Fig.~\ref{Fig2}b} and {\color{black}c}. Additionally, {\color{black}as indicated by arrows in Fig.~\ref{Fig2}i,j}, $I_\text{c}^\text{J}$ exhibits a specific symmetry: the behavior is invariant under the simultaneous reversal of the bias sweep direction (i.e. the sign of initial bias) and the sign of $H$. 

Specifically, within certain field ranges, $I_\mathrm{c-max}^\mathrm{J}$ is approximately four times larger than $I_\mathrm{c-min}^\mathrm{J}$ in {\color{black}Fig.~\ref{Fig2}g}. The large tunability is characterized by switching efficiency defined as $\eta= \frac{I_\mathrm{c-max}^\mathrm{J}-I_\mathrm{c-min}^\mathrm{J}}{I_\mathrm{c-max}^\mathrm{}+I_\mathrm{c-min}^\mathrm{J}}$, which approaches 60\% at $H=1.5$~Oe. Similar data for other devices (D4 and D5) are included in {\color{black}Supplementary.~1 and 2}. {\color{black} Figure~\ref{Fig2}h} summarizes $\eta$ across different devices, {\color{black}revealing a maximum $\eta$ of 65\%} at a minimal field of only 0.1~Oe.

~\\
\noindent\textbf{Electric switchable critical temperature}

\noindent
In {\color{black}Fig.~\ref{Fig2}k and l}, the $R-T$ profiles of D2, obtained after $\pm300~\mu$A pulse excitation, exhibit strong modulation, leading to a 30\% shift in the critical temperature defined at zero resistance ($T_\text{c0}$). The critical temperature control is analogous to SSV effects in FM/S/FM heterostructures, in which $T_\textit{c}$ suppression is governed by superconducting exchange interaction~\cite{You2002PRL,Blamire2017NM,Robinson2025NC}. Unlike conventional SSVs that require external $H$ as switching a knob,  
$T_\text{c}$ is manipulated by current pulse in our case, which is scarcely reported. This phenomenon represents another distinct feature of our junctions, offering a more facile and energy-efficient way for practical applications. As shown in {\color{black}Fig.~\ref{Fig2}l}, reversing $H_\text{z}$ to from -2.4~Oe to 2.4~Oe inverts the polarity of modulation. This response mirrors the behavior of $I_\mathrm{c}^\text{J}$, indicating that both effects originate from a common underlying mechanism. Similar data for D3 is presented in {\color{black}Supplementary Fig.~3}.

\begin{figure*}[thb]
	\begin{center}
		\includegraphics[width=0.8\textwidth]{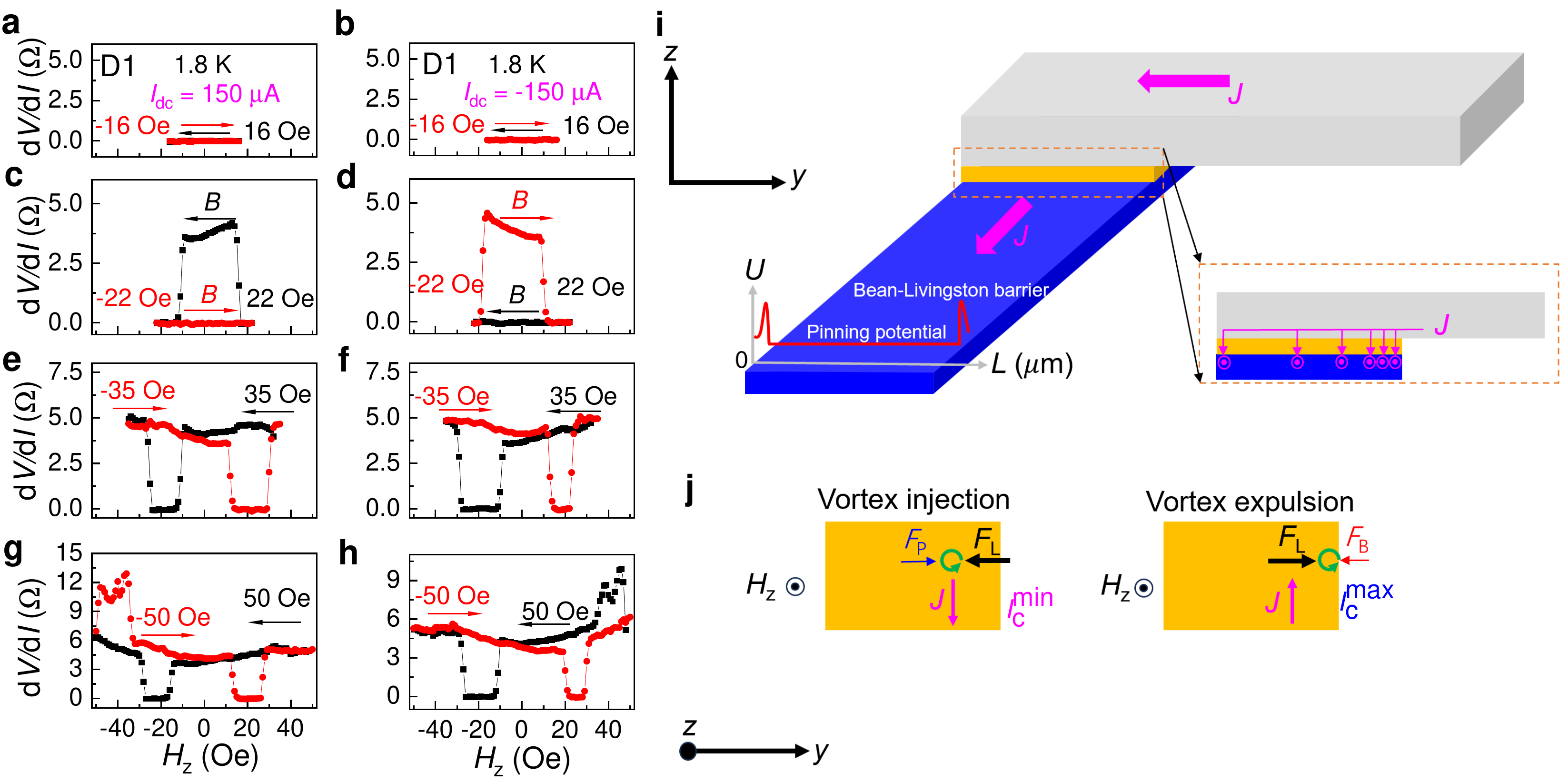}
	\end{center}
	\setlength{\abovecaptionskip}{-8 pt}
	\caption{(a) d$V$/d$I$ versus $I_\textrm{dc}$ curves for D1 at various $T$ and appropriate $P$ for $f$ = 6.48 GHz. (b) $V_\textrm{dc}$-$I_\textrm{dc}$ curve obtained by integrating the d$V$/d$I$-$I_\textrm{dc}$ curves at different temperatures in (a). $V_\textrm{dc}$ is normalized to the integer Shapiro step separation voltage $V_0 = hf/2e$. (c) d$V$/d$I$ as a function of normalized $V_\textrm{dc}$ at $T$ near $T_\mathrm{c}$.
	}
	\label{Fig3}
\end{figure*}

~\\
\noindent\textbf{Possible mechanisms}

\noindent
Before discussing the possible origins of the observed switchability, let's examine the transport characteristics in greater detail. {\color{black} Figure~\ref{Fig3}} presents $dV/dI$ measured by sweeping $H_\text{z}$ for D1, while keeping a constant bias of $\pm150~\mu$A. For a maximum sweeping field of $H_\text{sp-max} = 16$Oe in Fig~\ref{Fig3}a, b, the device is in the SC state with no hysteresis in $dV/dI$ curves . While at higher $H_\text{sp-max}$, complex hysteresis emerges, characterized by abrupt jumps between zero and finite resistance. Intriguingly, the hysteresis display distinct characteristics in different field ranges. 
For $H_\text{sp-max} = 21$Oe, the hysteresis loop reverses its direction when switching the sign of bias, as seen in {\color{black}Fig.~\ref{Fig3}c,d}. In contrast, at higher $H_\text{sp-max}$, the loops become independent of the bias sign in  {\color{black}Fig.~\ref{Fig3}e-h}. This suggests a crossover from current-dominated to field-dominated regimes by increasing field, revealing an intricate interplay between current and field in the underlining dynamics.


One potential explanation for this hysteresis is the presence of magnetic domains~\cite{MiaoF2024NC,LeT2024CPB}.  Consequently, the switch of $I_\text{c}^\text{J}$ could be attributed to current-induced spin reconfiguration via current-spin interactions~\cite{Zutic2004RMP,Ralph2007JMMM,Manchon2019RMP}. However, given that both NbSe$_2$ and Nb are well-established conventional superconductors lacking magnetic order, the magnetic domain hypothesis is rendered less probable. Furthermore, the abrupt jumps observed in {\color{black} Fig.~\ref{Fig2} and \ref{Fig3}} is inconsistent with typical domain dynamics, which generally manifest as more gradual transitions. 

A more plausible explanation involves vortex dynamics, where the observed jumps are related to sudden entry of isolated vortex.
It's important to emphasize that in our overlap cross-bar junction, $H_\text{z}$ is applied perpendicular to the junction plane. This configuration differs from that employed in Josephson effects, wherein $H$ is aligned parallel.  Note that no switching behavior is observed when similar amount of parallel field is employed, as seen in {\color{black} Supplementary Fig.~S4}. In the perpendicular configuration, the effective field at the junction edges is significantly enhanced due to demagnetization effects{\color{black}(see Supplementary Discussion)~\cite{Osborn1945PR}, as depicted in Fig.~\ref{Fig3}i.} This leads to magnetic flux injection into the junction, which is governed by the competition between the Bean-Livingston surface barrier and the driving Lorentz force $\textbf{F}_\text{L} = \textbf{J}\times\mathbf{\Phi}_0$, where $\mathbf{\Phi}_0$ denotes the flux quantum vector and $\mathbf{J}$ is the current density~\cite{Larkin1979JLTP,Tinkham2004Book}. {\color{black}Subsequently, the trapped vortex is driven by the applying current, generating dissipation and resulting in the suppression of critical current at the junction.} Given the crossbar geometry, it is reasonable to expect a non-uniform distribution of $J$, which is intensified at one edge due to the current crowding effect, {\color{black}as illustrated in Fig.~\ref{Fig3}i}. This spatial inhomogeneity renders $F_\mathrm{L}$ asymmetric, being stronger at the edge with higher $J$. This kinetic asymmetry plays a critical role in vortex injection and repulsion processes seen in {\color{black}Fig.~\ref{Fig3}j}, leading to non-volatile modulation of $I_\text{c}
$.

Within this framework, the switching behavior in {\color{black} Fig.~\ref{Fig2}} can be elucidated as follows: A positive bias ($I_\text{dc} > I_\text{sh}$) drives the vortex into the junction (vortex injection) at the edge of higher $F_\text{L}$, resulting the suppression of $I_\mathrm{c}^\text{J}$.  Conversely, a negative bias ($I_\text{dc} < -I_\mathrm{sh}$) facilitates vortex expulsion at the same edge, thereby restoring the state of higher $I_\mathrm{c}^\text{J}$. This mechanism is symmetric upon the simultaneous inversion of $I_\text{dc}$ and $H_\text{z}$, in agreement with {\color{black} Fig.~\ref{Fig2}}. The sudden jumps by vary $H_\text{z}$ in Fig.~\ref{Fig2} and \ref{Fig3} are interpreted by isolated vortex injection and repulsion. Regarding the hysteresis crossover in {\color{black} Fig.~\ref{Fig3}}, the dominant vortex dynamics evolves from current-driven at low $H_\text{z}$ to field-driven described by Bean's model at higher $H_\text{z}$~\cite{Tinkham2004Book,BeanRMP1964}. Consequently, the current switchable $T_\mathrm{c0}$ arises from the presence or absence of thermally activated vortex driven by applying current, i.e vortex creep effect.

{\color{black}Typically, $I_\mathrm{c}$ is expected to decay with increasing $H$ due to the accumulation of vortices and the suppression of order parameter, which is inconsistent with the staggered non-monotonicity observed in {\color{black}Fig.~\ref{Fig2}i and j}. This non-monotonicity suggests certain cancellation effects among the accumulated vortices~\cite{Hyun1998PRB}, counter-intuitive to common wisdom of vortex dynamics. This phenomenon is potentially interpreted as follows: the inter-vortex repulsion induced by the additional vortex stabilizes the  trapped flux configuration, thereby requiring higher current to depin the vortices. Similar non-monotonic trend have been observed near the upper critical field in dense flux lattices~\cite{Berlincourt1961PRL}, which was interpreted by collective pinning~\cite{Larkin1979JLTP}. However, the vortices in our system are minimal, it's surprising to observe the  non-monotonicity in this regime, which calls for further understanding of the underlying vortex dynamics. 
}




~\\
\noindent\textbf{Nonvolatile control of superconducting states}

\begin{figure*}[thb]
	\begin{center}
		\includegraphics[width=0.7\textwidth]{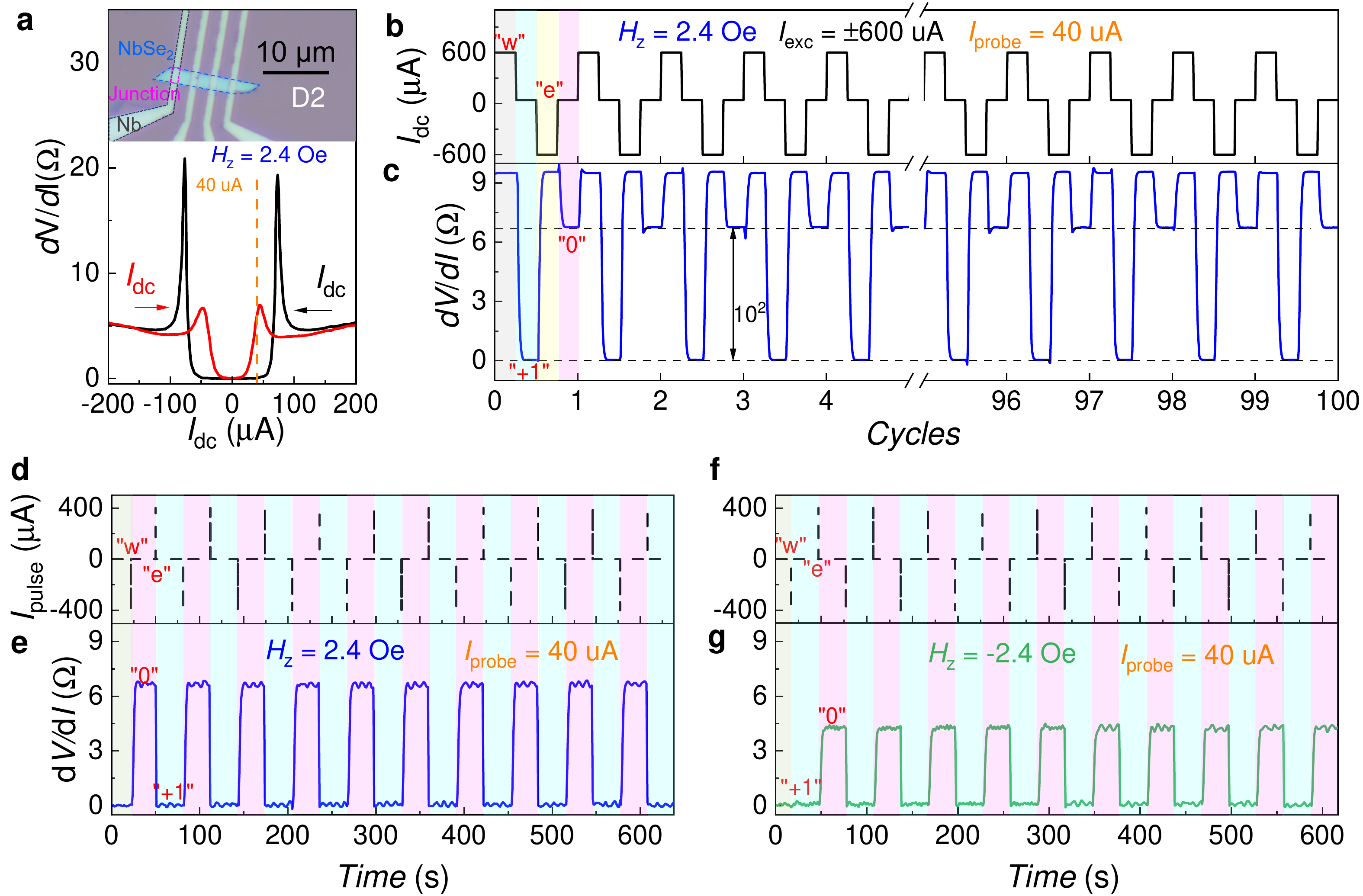}
	\end{center}
	\setlength{\abovecaptionskip}{-8 pt}
	\caption{\textbf{Nonvolatile control of the SC state in D2.} \textbf{a} d$V$/d$I$ versus $I_\textrm{dc}$ at $H_\textit{z}=2.4$~Oe. \textbf{c}  Switching characteristics between the superconducting and normal states monitored by d$V$/d$I$. The excitation current ($I_\textit{exc}$) is set to $\pm600~\mu$A. \textbf{d,e} Pulse-driven switching behavior excited by 50~$\mu$s current pulses of $\pm400~\mu$A  at  $H_\textit{z}=2.4$~Oe and -2.4~Oe, respectively. The probe current is set to 40~$\mu$A. Scale bar, 10 $\mu$m.		
	}
	\label{Fig4}
\end{figure*}

\noindent
In {\color{black}Fig.~\ref{Fig4}}, we demonstrates deterministic, nonvolatile control of SC states of D2 by exploiting the tunability of its $I_\text{c}^\text{J}$ . Switching between zero and finite resistance states, denoted as  "1" and "0" states, can be realized by setting the probe current between $I_\text{c-max}^\text{J}$ and  $I_\text{c-min}^\text{J}$. {\color{black}Figure~\ref{Fig4}a} presents the $dV/dI$ as a function of $I_\text{dc}$ at $H_\textit{z}=2.4$~Oe. 
In {\color{black}Fig.~\ref{Fig4}b, c}, reliable switching between "1" and "0" states is achieved by applying "write"-"read"-"erase"-"read" sequences with corresponding currents of -600~$\mu$A, 600~$\mu$A, and 40~$\mu$A, respectively. This switching behavior remains stable for over 100 cycles, underscoring the robustness of the device. 


The switching characteristics were further proved by applying a train of current pulses ($I_\text{p} = \pm400~\mu$A) to the device, each with a pulse width of 50~$\mu$s (instrumentation limited). {\color{black} This yields a switching energy of 1 pJ, which represents an experimental upper bound. It will be discussed below  that the ideal switching  energy is  significantly lower.}  As shown in {\color{black}Fig.~\ref{Fig4}d,e} (measured at $H_\text{z} = 2.4$~Oe), a positive pulse triggers a zero resistance state ("1"), which is subsequently switched back to finite ("0") by a negative pulse. In {\color{black}Fig.~\ref{Fig4}f,g}, the switching behavior  reverses upon field reversal. {\color{black}With a probe current of 40~$\mu$A, the "1"-state resistance is in the noise level, while the "0"-state resistance is approximately 7~$\Omega$, resulting in an lower bound on/off ratio of 100. This ratio is heavily limited by the resolution of our measurement setup. In principle, the superiority of switchable SC devices is reflected in the infinite on/off ratio, owing to the dissipationless supercurrent flow.}

~\\
\noindent{\Large\textbf{Discussion}}

\noindent
Below, let's estimate the intrinsic performance limits of our SC junctions. For a junction of width $w$, the transit time required for an Abrikosov vortex to traverse the junction is given by $t=w/v$. The vortex drift velocity, $v$, is determined by the balance between $F_\text{L}$ and the viscous drag, expressed 
as $v=\Phi_0 J/\eta$, wherein the viscosity is $\eta=B_\text{c2}\Phi_0/\rho_\text{n}$. Consequently, one yields $v=J\rho_\text{n}/B_\text{c}$. Given the typical parameters for  NbSe$_2$ (the upper critical field $B_\text{c2}\approx5$~T and normal state resistivity $\rho_\text{n} \approx 10~\mathrm{\mu\Omega.cm}$) and  a switching current density of  $J_\text{sh}\approx5\times10^5$~A/cm$^2$, $v$ is estimated as 100~m/s. Given $w\approx1\mathrm{\mu m}$, $t$ amounts to 10~ns. 
Furthermore, the minimum energy required to manipulate a single vortex corresponds to the work done by $F_\text{L}$ to drive the vortex to across the junction,  $E=I\Phi_0$.  This energy is exceptionally low ($<10^{-18} J$), satisfying the stringent power efficiency requirements for a cryogenic random access memory~\cite{Krasnov2015NC,Ortlepp2014IEEE}.


In summary, we report the robust electric switchability of critical current in crossbar SC junctions, achieved with significantly reduced activation currents and a substantial switching window. We also observe a non-volatile, electric modulation of critical temperature, which is an exotic feature in SC junctions. These effects are interpreted as a consequence of complex, asymmetric vortex dynamics inherent to the unique crossbar architecture.
These switchable crossbar junctions emerge as highly scalable active building blocks, hosting potential in advanced, energy efficient SC computation and brain-inspired SC neuromorphic architectures.

~\\


\noindent{\Large\textbf{Methods}}

~\\
\noindent\textbf{Device Fabrication} 
 
\noindent
The NbSe$_{2}$/Au/Nb crossbar junctions  were fabricated through the following sequential processes: First, NbSe$_{2}$ nanoflakes ($\sim$ 15 nm) were exfoliated from bulk crystals using blue tape and transferred onto a 300 nm-SiO$_{2}$/Si substrate. Next, electron beam lithography was employed for electrode patterning to define 1 $\mu$m-wide Au/Nb stripes. Finally,  6.5 nm-thick Au and 80 nm-thick Nb layers were sequentially deposited via DC magnetron sputtering under a base pressure of of 1$\times$ 10$^{-6}$ Torr.

~\\
\noindent\textbf{Transport measurements}  
	
\noindent	
Electrical transport measurements of the junctions were performed by a three-terminal method in a Physical Property Measurement System (PPMS, Quantum Design).A lock-in amplifier (SR830, Stanford Research) coupled with a 100 k$\Omega$ buffer resistor was employed to offer a small ac current ($I_\textrm{ac}$) for the detection of differential resistance (d$V$/d$I$). The current bias ($I_\textrm{dc}$) was supplied by a current source meter (Keithley 2450). Another current source meter (Keithley 2400) served as the power supply for the superconducting magnet, enabling precise control of the magnetic field at sub-gauss levels. $H_\text{z}$ was applied perpendicular to the junction plane. 

\noindent

~\\

~\\


~\\




~\\


~\\

\end{document}